# Ubermag: Towards more effective micromagnetic workflows

Marijan Beg[1,2], Martin Lang[2], and Hans Fangohr[2,3,4]

[1]Department of Earth Science and Engineering, Imperial College London, London SW7 2AZ, United Kingdom
[2]Faculty of Engineering and Physical Sciences, University of Southampton, Southampton SO17 1BJ, United Kingdom
[3]Max Planck Institute for the Structure and Dynamics of Matter, Luruper Chaussee 149, 22761 Hamburg, Germany
[4]Center for Free-Electron Laser Science, Luruper Chaussee 149, 22761 Hamburg, Germany

**Computational micromagnetics has become an essential tool in academia and industry to support fundamental research and the design and development of devices. Consequently, computational micromagnetics is widely used in the community, and the fraction of time researchers spend performing computational studies is growing. We focus on reducing this time by improving the interface between the numerical simulation and the researcher. We have designed and developed a human-centred research environment called Ubermag. With Ubermag, scientists can control an existing micromagnetic simulation package, such as OOMMF, from Jupyter notebooks. The complete simulation workflow, including definition, execution, and data analysis of simulation runs, can be performed within the same notebook environment. Numerical libraries, co-developed by the computational and data science community, can immediately be used for micromagnetic data analysis within this Python-based environment. By design, it is possible to extend Ubermag to drive other micromagnetic packages from the same environment.**

## I. Introduction

### A. Historical context

Computational micromagnetics has enjoyed increasing popularity since simulation codes such as OOMMF [1] have become available. As of early 2021, more than 3000 scientific manuscripts have been published[1] that refer to the OOMMF simulation software, presumably using it to enhance or enable their study.

The challenge of numerical micromagnetics' computational complexity, involving non-linear stiff partial differential equations with long-range interactions, has been addressed through new simulation methods that reflect the computing landscape's development. Magpar [2] pioneered the use of MPI to make use of high-performance computing clusters, Nmag [3] combined MPI with automatic code generation and performance optimised run-time compilation, and mumax$^3$ [4] exploits graphics processing unit (GPU) hardware to accelerate the numerical calculations.

In this work, we do not attempt to reduce the time numerical calculations take but instead focus on minimising the effort researchers need to invest in when setting up, driving, and analysing micromagnetic simulations to make their research more effective. We present Ubermag [5] – a micromagnetic simulation environment that makes micromagnetic simulations more flexible and effective. We review the conventional computational workflows and show how adopting Ubermag can help to make them more effective.

[1]https://math.nist.gov/oommf/oommf_cites.html

### B. Computational workflows in micromagnetics

The general workflow of using micromagnetic simulations is identical to many areas of computational science and consists of the following steps:

1) Decide what problem needs to be solved.
2) Translate this physics-based problem into the syntax understood by the simulation tool (often a configuration or script file).
3) Run simulation to compute results and write data files.
4) Analyse and visualise data files to obtain tables and plots.
5) Summarise and write up results and insights in a scientific manuscript or technical report. In this process, steps 1 to 4 may be repeated iteratively many times.

For simplicity, this generic summary ignores that some of the steps can also be done for some simulation packages through a graphical user interface (GUI).

Open-source micromagnetic simulation packages are written in different languages, often combining one language for the computational core, which solves the numerical problem, and a different language or syntax for the problem definition and user interface. As long as researchers do not need to change a simulation package's functionality, they can ignore the computational core languages. However, they need to know the syntax of configuration files or scripts from which the simulation is driven – workflow Step 2 requires the translation of the micromagnetic problem into the simulation tool syntax. For OOMMF [1], a Tcl-based syntax is used for configuration files, for mumax$^3$ [4] the Go language, and in Fidimag [6] micromagnetic simulations are configured with Python.

For post-processing, data analysis, and visualisation in workflow Step 4, researchers may be able to use relevant tools coming with the simulation package. However, due to the nature of research and the need to investigate new ideas, they may have to write additional data analysis and visualisation scripts. This requires understanding data file formats and conventions used by the simulation tool. Often, these scripts are not publicly available and simulation-package-specific – they process results produced by a particular simulation package and cannot be easily applied to results from other packages.

In workflow Step 5, one may want to run many simulations in a loop to explore parameter space (e.g. by changing geometry and material parameters) and gain a good understanding of the problem. In general, this requires writing additional code outside the simulation configuration files to generate simulation configuration files, and subsequently, additional post-processing and data analysis scripts.

For the reproducibility of scientific studies [7], it is required to be able to repeat each of the many simulations in the future. Ideally, one keeps a detailed log of all steps taken to obtain a particular table or figure, starting from the configuration file and execution parameters, including all post-processing, analysis, and visualisation steps.

A researcher may need to use more than one simulation tool because different micromagnetic simulation packages have different functionalities. Thus, the researcher must comprehend their capabilities, installation, configuration, user interfaces, and data formats.

The combined effort of these steps establishes the learning curve for users of micromagnetic software. For computationally-minded scientists, this is probably faster to master than for those with expertise in other fields such as experiments, pure theory, or device design. As simulation use becomes more widespread, there are also more users with less computational training. We developed Ubermag[2] [5] to make computational micromagnetic workflows and studies more effective from usability and user-centred perspective [8], [9].

## II. UBERMAG

### A. Introduction

Ubermag [5] can be understood as a Python-based layer that lies on top (German *über*) of existing micromagnetic simulation tools, exposing them to Python's ecosystem and integrating them into the Jupyter environment. We decided to start with OOMMF [1] as the micromagnetic calculator to carry out the numerical simulations, but work is underway to make mumax[3] [4] available as a first alternative calculator. This section shows an example of Ubermag in practice, and we discuss some of the design choices in section III.

### B. Defining the micromagnetic problem (`micromagneticmodel`)

The first step in a computational micromagnetic workflow is to decide what micromagnetic problem to solve (workflow Step 1 in Sec. I-B). The `micromagneticmodel` [10] library provides a Python-based domain-specific language for defining micromagnetic systems (which could also be called problems or models). It is a description of the physics we are interested in. This machine-readable description is not concerned with finding a solution of the problem: the actual numerical solving is discussed in Sec. II-C. To fully define a micromagnetic system, the following components must be specified: (i) energy equation – the sum of energy density terms, (ii) dynamics equation – the sum of dynamics terms governing the magnetisation dynamics, and (iii) magnetisation field, which uniquely defines the state of the system.

For example, let us define a micromagnetic system modelling a thin-film Permalloy square sample with $L = 100\,\text{nm}$ edge length and $5\,\text{nm}$ thickness. Its energy equation consists of

[2]https://ubermag.github.io/

ferromagnetic exchange, Zeeman, and demagnetisation energy terms:

$$E = \int_V \left[ -A\mathbf{m} \cdot \nabla^2 \mathbf{m} - \mu_0 M_s \mathbf{m} \cdot \mathbf{H} + w_\text{d} \right] \text{d}V, \quad (1)$$

where $A = 13\,\text{pJ}\,\text{m}^{-1}$ is the exchange energy constant, $M_s = 8 \times 10^5\,\text{A}\,\text{m}^{-1}$ magnetisation saturation, $w_\text{d}$ demagnetisation energy density, $\mathbf{H}$ an external magnetic field, and $\mathbf{m} = \mathbf{M}/M_s$ the normalised magnetisation field. The magnetisation dynamics is governed by the Landau-Lifshitz-Gilbert equation's [11] precession and damping terms:

$$\frac{\partial \mathbf{m}}{\partial t} = -\frac{\gamma_0}{1+\alpha^2} \mathbf{m} \times \mathbf{H}_\text{eff} - \frac{\gamma_0 \alpha}{1+\alpha^2} \mathbf{m} \times (\mathbf{m} \times \mathbf{H}_\text{eff}), \quad (2)$$

where $\gamma_0 = 2.211 \times 10^5\,\text{m}\,\text{A}^{-1}\,\text{s}^{-1}$. Although we simulate a Permalloy sample, we use an artificially large value for Gilbert damping $\alpha = 0.2$ to simplify the magnetisation dynamics in our example. The initial magnetisation state is a vortex defined for each point $\mathbf{r} = (x, y, z)$ in the sample as $\mathbf{m}(\mathbf{r}) = (m_x, m_y, m_z) = (-cy, cx, 0.1)/\sqrt{c^2 y^2 + c^2 x^2 + (0.1)^2}$, with $c = 10^9\,\text{m}^{-1}$. The normalisation to $|\mathbf{m}| = 1$ is done by Ubermag automatically. The Python code for defining the micromagnetic system in Ubermag is:

```python
import discretisedfield as df
import micromagneticmodel as mm

# Geometry
L = 100e-9  # sample edge length (m)
thickness = 5e-9  # sample thickness (m)

# Discretisation cell (lengths in m)
cell = (5e-9, 5e-9, 5e-9)

# Material parameters (Permalloy)
Ms = 8e5  # saturation magnetisation (A/m)
A = 13e-12  # exchange energy constant (J/m)

# Dynamics (LLG equation) parameters
gamma0 = 2.211e5  # gyromagnetic ratio (m/As)
alpha = 0.2  # Gilbert damping

system = mm.System(name='vortex_dynamics')

# Energy equation - we omit Zeeman energy term,
# because H=0
system.energy = mm.Exchange(A=A) + mm.Demag()

# Dynamics equation
system.dynamics = (mm.Precession(gamma0=gamma0) +
                   mm.Damping(alpha=alpha))

# Initial magnetisation state
def m_init(point):
    x, y, z = point
    c = 1e9  # (1/m)
    return (-c*y, c*x, 0.1)

# Sample's centre is placed at origin
region = df.Region(p1=(-L/2, -L/2, -thickness/2),
                   p2=(L/2, L/2, thickness/2))
mesh = df.Mesh(region=region, cell=cell)
system.m = df.Field(mesh, dim=3,
                    value=m_init, norm=Ms)
```

At this point, we have a system object modelling the problem we want to simulate on a finite-difference mesh. So far, we have no means to solve the associated equations – we have only expressed the relevant physics in a computer-readable way. We have completed workflow Step 2 in Sec. I-B.

### C. Using a micromagnetic calculator (`oommfc`)

To solve the equations numerically (workflow Step 3 in Sec. I-B), we use OOMMF. In the framework of Ubermag, OOMMF is one (of potentially many) *micromagnetic calculators*. The corresponding library is called `oommfc`, *i.e.* OOMMF calculator. We call the methods to change the magnetisation field *drivers*, imagining that they drive the system through different points in phase space either by minimising the energy Eq. (1) or solving the dynamics Eq. (2).

In this example, we are going to use the energy minimisation driver from `oommfc` library to act on our defined system:

```
import oommfc as mc   # micromagnetic calculator
md = mc.MinDriver()
md.drive(system)
```

The `md.drive(system)` call will change the state of the system so that it corresponds to the nearest energy minimum. Ubermag translates the micromagnetic model `system` to a configuration file and then calls the OOMMF executable. When OOMMF has completed the calculation, Ubermag reads output files with spatially-resolved and tabular data. The obtained magnetisation field is used to update the state of the `system` object. In our example, we are minimising the system's energy which results in a vortex magnetisation field, as we show in Fig. 1.

Let us now apply an external magnetic field $H = 10^4 \, \text{Am}^{-1}$ in the positive $x$-direction to displace the vortex core. We can achieve that by adding the Zeeman energy term to the energy equation and driving the system again:

```
H = (1e4, 0, 0)   # an external magnetic field (A/m)
system.energy += mm.Zeeman(H=H)
md.drive(system)
```

We have now moved the state of the system in phase space by displacing the vortex core, as we show in Fig. 2, cell 13. Finally, we set the external magnetic field to zero, simulate the time-evolution of the magnetisation field for $5\,\text{ns}$, and save the output in $500$ steps (*i.e.* save once every $10\,\text{ps}$):

```
system.energy.zeeman.H = (0, 0, 0)
td = mc.TimeDriver()
td.drive(system, t=5e-9, n=500)
```

If the commands described above are used in a Jupyter Notebook (see Sec. III-A), they can be executed interactively, modified and re-executed as often as desired within the same notebook. This enables the researcher to drive and modify the system iteratively to probe and understand its behaviour.

### D. Data analysis and visualisation

After driving the system, we can analyse and visualise data (workflow Step 4 in Sec. I-B). There are three Ubermag libraries that can be used for data analysis and visualisation: `discretisedfield`, `ubermagtable`, and `micromagneticdata`.

The purpose of `discretisedfield` is to represent spatially-resolved vector and scalar fields as `numpy` arrays,

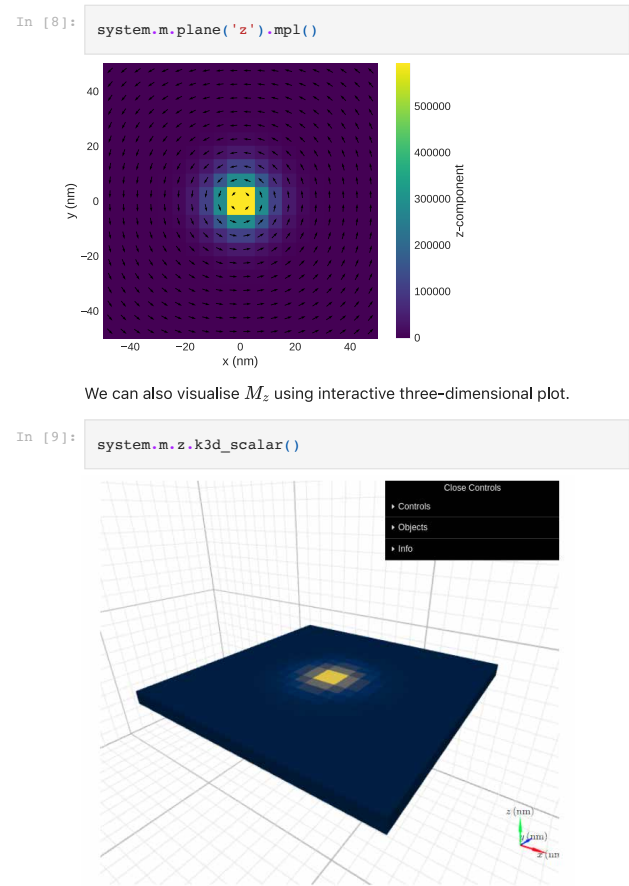

Fig. 1. Magnetisation field visualisation using `matplotlib` (Cell 8) and `k3d` (Cell 9) in Jupyter environment.

making all linear algebra and array manipulation methods available, and to provide visualisation capabilities for spatially resolved data using `matplotlib` for two-dimensional and `k3d` for interactive three-dimensional plots. Fig. 1 shows two- and three-dimensional plots from a Jupyter notebook that also contains the code snippets shown above.

The `ubermagtable` library uses `pandas` dataframes for representing time- or step-dependent scalar data, its manipulation, and visualisation, as we show an example in cells 10 and 11 in Fig. 2. In order to analyse different results from subsequent calls of the `drive` method from time- and minimisation-drivers, `micromagneticdata` is used. We show an example in cell 12 in Fig. 2.

All analysis libraries allow building interactive plots, consisting of GUI widgets for varying different parameters and interactive visualisation. In cell 13 in Fig. 2, we show an example where the step slider can be moved, and the displayed magnetisation vector field will update automatically. Interactive plotting environments enable users to inspect and explore the data and gain additional insight.

A Jupyter notebook containing the entire example presented in this work is available in the repository [12] accompanying this work.

## III. Discussion

### A. Jupyter Notebook

We have chosen the Jupyter Notebook [13], [14] as the primary user interface because "Jupyter helps humans to think and tell stories with code and data" [9], and has been designed for that purpose by researchers for researchers. A notebook can contain narrative (human-readable text, equations, and figures), code, and code's output in a single document. Apart from using it to interactively explore and understand simulation results, we can include an entire micromagnetic study in a single self-consistent document. Jupyter notebooks are powerful environments for data analysis [15].

Jupyter notebooks make research more reproducible [16] and help share studies either as a read-only HTML or PDF or as executable documents. Services such as myBinder [17] make it possible to execute notebooks in the cloud within a web browser without installing any software locally. Software required to run the notebook can be defined in a separate file which, together with notebooks, can be made available in a public repository. Some examples include works in Refs. [18], [19], [12].

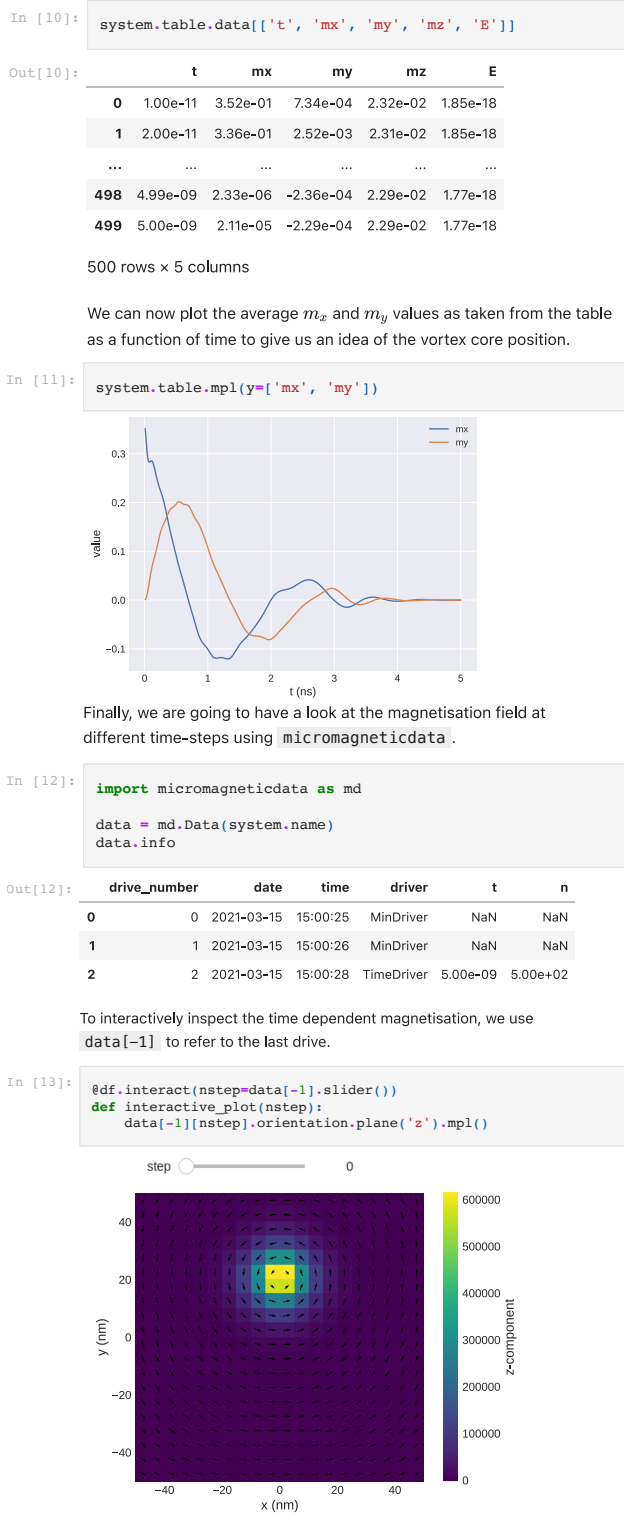

Fig. 2. Scalar data table representation (Cell 10) and its visualisation (Cell 11) with `ubermagtable`, accessing micromagnetic data from subsequent drives (Cell 12), and building a custom interactive plot in the Jupyter environment (Cell 13).

### B. Exploiting existing scientific software ecosystem

There are multiple reasons why we have chosen Python as the programming language used in Ubermag. This includes the active science user-community and a wide range of existing scientific libraries for data analysis and visualisation such as `numpy` for linear algebra, `scipy` for numerical analysis, `pandas` for tabular data analysis and visualisation, `matplotlib` for visualisation, `k3d` for interactive three-dimensional visualisation, `scikit-learn` for machine learning, etc. Python is a language that is relatively accessible to scientists and engineers [20], and as such, adding customised research code to a study is more easily achievable. A use-case for such a situation is the computation of custom mathematical operations on discretised fields. For example, one can compute this expression for the winding number:

$$S = \frac{1}{4\pi} \iint \mathbf{m} \cdot \left( \frac{\partial \mathbf{m}}{\partial x} \times \frac{\partial \mathbf{m}}{\partial y} \right) \mathrm{d}x\mathrm{d}y \qquad (3)$$

in `discretisedfield` using the following notation, which exploits operator overloading:

```
import math
m = system.m.orientation.plane('z')
q = m @ (m.deriative('x') & m.derivative('y'))
S = 1/(4*math.pi) * df.integral(q * df.dx*df.dy)
```

and modify it further, using all the power of Python and its supporting libraries, as required by the research task.

Finally, Python (as any general-purpose programming language) provides execution flow control commands, such as loops and conditional branching, to enable computational workflow automation without leaving the computational environment. This is important for workflow Step 5 in Sec. I-B.

## C. Open Science

All Ubermag libraries are written in Python and available as open source, making contributions by the community possible. We welcome all contributions, including improvements of documentation and tutorials[3]. As Ubermag has been designed with this in mind, it is possible to integrate other finite-difference based micromagnetic calculators into the package – for mumax[3] this work has started. Multiple micromagnetic calculators will open up new opportunities: by changing one line of code, the researcher can repeat the same micromagnetic study with a different micromagnetic calculator. Similarly, different calculators could be used subsequently within the same notebook session.

## D. Installation

Ubermag can be installed on all major operating systems via `pip` and `conda` package managers. Installing the Ubermag meta-package [5] installs all its dependencies, including OOMMF. For reproducibility purposes, the availability of past versions of Ubermag in public `pip` and `conda` clouds is not guaranteed. Therefore, to permanently retain the working copy of the computing environment, a Docker[4] image can be built and made available in a public repository with an assigned digital object identifier (DOI). While packaging the Python-based Ubermag libraries is a reasonable effort, conda packaging of third-party software such as OOMMF, and potentially mumax[3] and others in the future is likely to be not sustainable for the Ubermag project team: users would thus need to install OOMMF or mumax[3] themselves, before installing Ubermag.

## IV. Conclusion

Ubermag provides a Jupyter-Notebook-based interface to micromagnetic simulations. The user defines micromagnetic problems in a computer-readable way and solves them using external calculators such as OOMMF. Results can be explored, annotated, and refined interactively and iteratively. Fully automatic parameter space exploration can be realised within the same environment. In practical terms, this holds the potential to enable researchers to work more effectively.

In conceptual terms, Ubermag introduces a machine-readable definition of a micromagnetic problem and an abstraction layer above existing simulation packages. The problem's numerical solution is delegated by automatic translation of the problem definition into the tool-specific configuration syntax and subsequent execution of the tool.

Related projects in other scientific fields are the Atomic Simulation Environment [21], and the self-documenting data standard NeXuS [22] for neutron, x-ray and muon science.

## V. Acknowledgments

This work was financially supported by the Horizon 2020 European Research Infrastructure OpenDreamKit project (676541) and the EPSRC Programme grant on Skyrmionics (EP/N032128/1).

---

[3] https://ubermag.readthedocs.io/
[4] https://www.docker.com/